# Cosmic-ray exposure ages of fossil micrometeorites from mid-Ordovician sediments at Lynna River, Russia.

M. M. M. Meier[a]*, B. Schmitz[b], A. Lindskog[a], C. Maden[c] and R. Wieler[c]

[a]Lund University, Department of Geology, Sölvegatan 12, SE-22362 Lund, Sweden
[b]Lund University, Department of Physics, SE-22100 Lund, Sweden
[c]ETH Zurich, Department of Earth Sciences, CH-8092 Zurich, Switzerland

*Corresponding author: matthias.meier@geol.lu.se;

**Abstract:** We measured the He and Ne concentrations of 50 individual extraterrestrial chromite grains recovered from mid-Ordovician (lower Darriwilian) sediments from the Lynna River section near St. Petersburg, Russia. High concentrations of solar wind-like He and Ne found in most grains indicate that they were delivered to Earth as micrometeoritic dust, while their abundance, stratigraphic position and major element composition indicate an origin related to the L chondrite parent body (LCPB) break-up event, 470 Ma ago. Compared to sediment-dispersed extraterrestrial chromite (SEC) grains extracted from coeval sediments at other localities, the grains from Lynna River are both highly concentrated and well preserved. As in previous work, in most grains from Lynna River, high concentrations of solar wind-derived He and Ne impede a clear quantification of cosmic-ray produced He and Ne. However, we have found several SEC grains poor in solar wind Ne, showing a resolvable contribution of cosmogenic $^{21}$Ne. This makes it possible, for the first time, to determine robust cosmic-ray exposure (CRE) ages in these fossil micrometeorites, on the order of a few hundred-thousand years. These ages are similar to the CRE ages measured in chromite grains from cm-sized fossil meteorites recovered from coeval sediments in Sweden. As the CRE ages are shorter than the orbital decay time of grains of this size by Poynting–Robertson drag, this suggests that the grains were delivered to Earth through direct injection into an orbital resonance. We demonstrate how CRE ages of fossil micrometeorites can be used, in principle, to determine sedimentation rates, and to correlate the sediments at Lynna River with the fossil meteorite-bearing sediment layers in Sweden. In some grains with high concentrations of solar wind Ne, we nevertheless find a well-resolved cosmogenic $^{21}$Ne signal. These grains must have been exposed for up to several 10 Ma in the regolith layer of the pre-break-up L chondrite parent body. This confirms an earlier suggestion that such regolith grains should be abundant in sediments deposited shortly after the break-up of the LCPB asteroid.

# 1. INTRODUCTION

## 1.1. Traces of the L chondrite parent body break-up

Traces of the disruption of the L chondrite parent body (LCPB) asteroid 470 Ma ago have been preserved in both extraterrestrial and terrestrial records. In a majority of the L chondrites falling on Earth today, shock features such as fracturing, remelting and shock-darkening are abundant (e.g., Stöffler et al., 1992; Rubin, 1994). The loss of radiogenic noble gases ($^4$He and $^{40}$Ar) in these meteorites can be used to date the age of the gas-loss event to about 0.5 Ga ago (Anders, 1964; Bogard, 1995; Haack et al., 1996), with the most recent Ar–Ar ages converging on 470 ± 6 Ma (Korochantseva et al., 2007; Weirich et al., 2012). Condensed (i.e., slowly deposited) marine sediments formed on Earth around this time, in the Middle Ordovician time-period, record a sudden increase in the flux of extraterrestrial material, ranging from sub-millimeter-sized extraterrestrial dust grains found in coeval sediments from Sweden (Schmitz et al., 2003), China (Cronholm and Schmitz, 2010; Alwmark et al., 2012) and Russia (Korochantsev et al., 2009; Lindskog et al., 2012), to centimeter-sized fossil meteorites found so far only at different localities in Sweden (Thorslund and Wickman, 1981; Nyström et al., 1988; Tassinari et al., 2004). Also, an increase in the number of craters formed within a few tens of Ma after the LCPB asteroid disruption event has been suggested (e.g., Schmitz et al., 2001; Korochantseva et al., 2007). For the mid-Ordovician Lockne crater, it has been demonstrated that the impactor likely was an L chondrite (Alwmark and Schmitz, 2007; Schmitz et al., 2011). The mineral that carries the decisive evidence of extraterrestrial origin in all these cases is chromite (ideal formula: $FeCr_2O_4$), which has proven to be extraordinarily resilient to terrestrial weathering and diagenesis over geologic timescales (Schmitz et al., 2001). The extraterrestrial origin of chromite can be determined from its characteristic (L chondritic) composition in Cr, Fe, Al, Mg, Ti, V, Mn, Zn abundances (Bunch et al., 1967; Schmitz et al., 2001), the O isotopic composition (Greenwood et al., 2007; Heck et al., 2010) and the presence of cosmic-ray produced, or occasionally solar wind-derived, He and Ne (Heck et al., 2004; 2008; Meier et al., 2010; see next paragraph). A recent review of the topic of extraterrestrial chromite in terrestrial sediments is given by Schmitz (2013).

## 1.2. Earlier noble gas work on extraterrestrial chromite grains

Heck et al. (2004) measured the He and Ne concentrations of (batches of) chromite grains recovered from nine different fossil meteorites found in mid-Ordovician limestone from a quarry in southern Sweden. The isotopic composition of Ne in these grains is dominated by a $^{21}$Ne-rich component, produced during transfer of the meteorite from the asteroid belt to Earth by spallation of target atoms by highly energetic cosmic-ray particles. The cosmic-ray exposure (CRE) ages calculated from the concentration of this "cosmogenic" Ne, and an empirically determined production rate for chromite target chemistry, range from 0.1 to 1 Ma, increasing upward with stratigraphic position. This confirms that the fossil meteorites started exposure to cosmic-rays at the same moment, but spent different amounts of time in space before crossing paths with Earth. The CRE ages of these fossil meteorites are much shorter than the range of 3–60 Ma (Marti and Graf, 1992) observed in recently fallen L chondrites, which suggests an origin in an event that produced a great deal of of debris injected directly into resonant orbits (Nesvorný et al., 2009). The difference in CRE age of about 0.9 Ma between the lowermost and the uppermost fossil meteorites analyzed, and their vertical distance of ~3.4 m in the sediment yields an average sedimentation rate over that interval of ~3.8 mm/ka (Figure 1), in accord with bio-stratigraphic estimates (e.g., Schmitz et al., 1996).
Schmitz et al. (2003) found extraterrestrial chromite grains dispersed in the sedimentary rocks con-taining the fossil meteorites, the abundance of which increased by two orders of magnitude at the base of the fossil-meteorite-bearing sedimentary layers. Their elemental composition (identical to L chondritic chromite) and their oxygen isotopes (Heck et al., 2010) confirm a connection to the LCPB break-up. Since these sediment-dispersed extraterrestrial chromite (SEC) grains are much more abundant – each kg of limestone within the fossil-meteorite-bearing layers at the Swedish localities contains about 3 SEC grains larger than 63 μm (Schmitz et al., 2003) – and therefore easier to find than fossil meteorites, the He, Ne analysis of SEC grains was originally envisioned to provide a means to determine sedimentation rates in SEC-rich sediments at very high temporal resolution, and thus to provide a precise, multi-million-year monitor of the extraterrestrial influx of material to Earth in the aftermath of an asteroidal break-up. However, when Heck et al. (2008) measured He and Ne in the first batches of SEC grains, they found that the grains contained solar-wind-derived He and Ne in concentrations so high that the cosmogenic signal was not resolvable. A study of He and Ne in *individual* SEC grains (Meier et al., 2010) showed that the vast majority of them contain solar wind He and Ne. Since solar wind ions penetrate less than ~100 nm into exposed surfaces, the SEC grains must thus have been exposed to the solar wind on an individual basis, i.e., the solar-wind-rich SEC grains had to derive from dust-sized grains produced in the LCPB disruption event. An alternative explanation, where the SEC grains derive from weathered L chondritic regolith breccia meteorites, was dismissed because the abundance of such meteorites among the L chondrites is far too low (~3%; Bischoff and Schultz, 2004) to explain the high abundance of solar wind-rich grains among the SEC grains. Meier et al. (2010) also found that many individual grains indicated a contribution of cosmogenic Ne, which corresponds, in some cases, to a CRE age of several tens of millions of years. This cannot be explained with the <1 Ma transit times from the asteroid belt, required both from the orbital decay times of micrometeorite orbits induced by Poynting–Robertson-drag (e.g., Burns et al., 1979; but see also Klačka et al., 2014) and the grains' position relative to the LCPB break-up event in the sedimentary succession. Instead, Meier et al. (2010) suggested that these pre-exposed grains must have resided for a total of several tens of millions of years in the topmost few meters of the regolith layer of the pre-break-up LCPB asteroid. The analysis of more SEC grains from Sweden and China showed that the deposition of large numbers of solar-wind-rich grains at 470 Ma is a global, stratigraphic event (Alwmark et al., 2012). In the present work, we study the He and Ne concentration of SEC grains from yet another mid-Ordovician site, the Lynna River section in Russia. The SEC grains at Lynna River are highly concentrated (up to ~10 grains/kg) and well preserved compared to other sites (Lindskog et al., 2012; see next paragraph), and therefore offer a more pristine record of the LCPB debris falling to Earth 470 Ma ago. Preliminary results from this study have been presented in Meier et al. (2013a).

**1.3. The mid-Ordovician sedimentary rocks at Lynna River**

The Lynna River section is located just southwest of the village of Kolchanovo, south of Lake Ladoga, ca. 120 km east of St. Petersburg. The outcrop, overlooking the northern bank of the Lynna River just before it discharges into the larger Syas River, hosts ca. 10 m of Middle Ordovician rocks, which are part of a laterally extensive sediment blanket that formed in an epicontinental sea covering large parts of the Baltoscandian region (including western Russia; Lindskog et al., 2012, and references therein). The local rocks span the (uppermost) middle Volkhov through middle Kunda Baltoscandian stages, lower Darriwilian global Stage; see Bergström et al. (2009) for a correlation to other regional stratigraphic schemes. A pilot search for SEC grains in this section was done by Korochantsev et al. (2009), who found 0.6 to 2.9 SEC grains per kg of limestone. Lindskog et al. (2012) extended this study to include much more material sampled over a larger stratigraphic range, and found up to 10 SEC grains per kg in some samples – around a factor of three more than typical concentrations in coeval rocks in Sweden (Schmitz et al., 2003; Häggström and Schmitz, 2007) and China (Cronholm and Schmitz, 2010). Furthermore, the grains proved to be better preserved than is typical at other known localities. Most SEC grains from Lynna River are very angular to sub-angular, indicating minimal transport in the terrestrial environment. Some Lynna River SEC grains show Ni-enrichment, sometimes taking the form of MgO/FeO/NiO-rich rims, which has not been observed at other localities (Korochantsev et al., 2009; Lindskog et al., 2012).

## 2. SAMPLES & METHODS

### 2.1. SEC grain preparation, identification and mass determination

The SEC grains analyzed for this study derive from samples Ly3 and Ly4 in the study of Lindskog et al. (2012). These beds have vertical extensions of approximately 15 cm and 5 cm, respectively, and are located, respectively, just below and above the tentative boundary between the lower-laying *Lenodus variabilis* and the *Yangtzeplacognathus crassus* conodont stratigraphic zones. These two zones have also been identified in Sweden (Schmitz et al., 2003) and China (Cronholm and Schmitz, 2010). In Sweden, the boundary between them is located within the fossil-meteorite-bearing layers, with chromite grains from the uppermost fossil meteorites in the *Lenodus variabilis* zone yield CRE ages of about 0.3 Ma, while the lowermost fossil meteorites in the *Yangtzeplacognathus crassus* zone yield CRE ages of 0.5–0.6 Ma (Heck et al., 2004; 2008). The Lynna River SEC grains were extracted by dissolving the limestone in 6 M hydrochloric and 3.8 M hydrofluoric acid to remove carbonates and silicates, respectively. After neutralization, the residues were sieved, and the 63–355 μm size fraction was searched for opaque grains under an optical microscope. Candidate chromite grains were analyzed using a scanning electron microscope (SEM; Hitachi S-3400N) with an attached energy-dispersive spectrometer (EDS; Oxford Instruments Inca X-Sight; Co-standard) on unpolished surfaces, for qualitative identification of extraterrestrial chromite. We excluded all grains that did not correspond (within measuring error) to the typical L chondritic chromite composition (MgO ~1.5–4 wt%, $Al_2O_3$ ~5–8 wt%, $TiO_2$ ~1.4–3.5 wt%, $V_2O_3$ ~0.6–0.9 wt%, $Cr_2O_3$ ~55–60 wt%, and FeO ~25–30 wt%; Cronholm and Schmitz, 2010, and references therein), in particular regarding the concentrations in $TiO_2$ and $V_2O_3$, which have previously been found to be most diagnostic of extraterrestrial origin (Cronholm and Schmitz, 2010). In total, 34 grains from Ly3, and 16 grains from Ly4 were selected for analysis. The mass of the grains was measured on a micro-balance with a typical accuracy of ~0.3 μg. Each grain's mass determination was bracketed by two tare measurements of the empty balance. The uncertainty of the individual grain mass corresponds to the difference between the two adjacent tare measurements (the balance was recalibrated after each tare-grain-tare cycle). For grains with resulting mass uncertainties larger than 30%, the mass was instead determined using the following empirical relationship between cross-sectional area in a SEM image (A), and grain volume (V):

$$V = \alpha \times A^{1.5}; \alpha = 0.5 \pm 0.15$$

This empirical relationship was previously determined with meteoritic chromite grains imaged both with SEM and later (in 3D) with synchrotron radiation X-ray tomographic microscopy. The typical uncertainty (1σ) of this approach is ±30%. The mass was then calculated from estimated volume and the density of chromite (~4.4 g/cm$^3$ for L chondritic major element composition). For most of the grains, the resulting masses are identical, within uncertainties, with the masses determined on the balance, confirming the viability of the approach.

## 2.2. Noble gas analysis

Grains were placed individually into the holes of an aluminum sample holder, which was then placed into an extraction line underneath a viewport, and pumped to ultra-high vacuum (~$10^{-10}$ mbar) for 48 h. The noble gas analysis was done with the same instruments and protocols as used earlier by Meier et al. (2010) and Alwmark et al. (2012), therefore we will only give an abridged description here. We use a low blank extraction line with $ZrO_2/TiO_2$ getters and three cold traps cooled with liquid $N_2$ to remove background gases such as $H_2$, $CH_4$, $H_2O$, Ar and $CO_2$. The sample gas is then measured using a compressor-source mass spectrometer, where a molecular drag pump concentrates the sample gas almost quantitatively into the ion source volume, thereby leading to an increase in sensitivity of the mass spectrometer by up to two orders of magnitude (Baur, 1999). The low electron acceleration voltage of 40 V leads to almost negligible interferences of doubly-charged $^{40}Ar$ and $^{44}\{CO_2\}$ on masses 20 and 22, respectively. Before the sample gas extraction, all species of interest (HD, $^3He$, $^4He$, $H_2^{16}O$, $^{20}Ne$, $^{21}Ne$, $^{22}Ne$, $^{40}Ar$, $^{44}\{CO_2\}$) are measured in four cycles in peak-jumping mode ($H_2^{16}O$ is measured to control the interference of $H_2^{18}O$ on mass 20). Then, the sample gas is extracted by heating the SEC grain to the point of melting or slow vaporization with a Nd:YAG IR-laser ($\lambda$ = 1064 nm) for about 30–60 seconds. After extraction, all nine species are measured again in seven more cycles. No valves are operated during the whole procedure. The signal for each species is then extrapolated forward from the four pre-extraction cycles, and backward from the seven post-extraction cycles, to the moment of extraction (beginning of laser firing). The difference between the two extrapolated values is then considered to be the sample signal. This approach is described in more detail in Heck et al. (2007). Spectrometer sensitivity was calibrated with known amounts of a mixture of pure He and Ne (Heber et al., 2009). Interference corrections on masses 20 and 22 never exceeded 1% to the measured signals of $^{20}Ne$ and $^{22}Ne$.

## 3. RESULTS

### 3.1. Masses, elemental composition, He, Ne ratios and concentrations

The He, Ne data of 50 SEC grains from Lynna River samples Ly3 (34) and Ly4 (16), as well as masses and the (qualitative) elemental composition are given in Table 1. The masses of the individual grains vary from 0.5 to 11.1 µg (corresponding to average diameters of 60 to 170 µm). Two grains from Ly3 (Ly3-Cr03 and Ly3-Cr79) were lost during lasering. The $^4He$ and $^{20}Ne$ concentrations of the 48 analyzed grains vary by about three orders of magnitude. The most gas-rich grains contain about $5 \times 10^{-2}$ cm$^3$ STP $^4He$/g, and about $4 \times 10^{-4}$ cm$^3$ STP $^{20}Ne$/g. While the $^{20}Ne$ concentration range is similar to SEC grains from other localities, the $^4He$ concentrations in the Lynna River grains are up to a factor of ~5 higher than the values observed in the previous studies at other localities (Meier et al., 2010; Alwmark et al., 2012). Since trapped He is more easily lost than Ne, the higher $^4He/^{20}Ne$ ratios confirm the conclusion of Lindskog et al. (2012) that the Lynna River grains are exceptionally well preserved. The $^4He$ and $^{20}Ne$ concentrations reported here are of the same order of magnitude as the respective values in recent micrometeorites from Antarctica (Osawa and Nagao, 2002; see also Figure 2) and Greenland (Olinger et al., 1990). The $^3He/^4He$ ratio in most SEC grains from Lynna River is very similar to the lowest ratios observed in lunar soil (~$2.1 \times 10^{-4}$; Benkert et al., 1993). It is also similar to values in recent micrometeorites from Antarctica (Osawa and Nagao, 2002), as shown in Figure 2, and in SEC grains from other localities (Meier et al., 2010). Seven grains have $^3He/^4He$ ratios that are significantly lower ($<1 \times 10^{-4}$), indicating a contribution from atmospheric or radiogenic He. In another seven grains, $^3He$ was below the detection limit, resulting in upper limits $<1 \times 10^{-5}$ for the $^3He/^4He$ ratio. In the Ne three-isotope-plot (Figure 3), most of the grains plot close to a line connecting the Ne isotopic composition of the solar wind (Heber et al., 2009) with the composition of the implantation-fractionated solar wind (formerly known as the solar energetic particles (SEP) component; Grimberg et al., 2006). The data points plot predominantly to the right side of the solar wind fractionation line, indicating a contribution of $^{21}$Ne-rich, cosmogenic Ne. This was observed before in SEC grains from other localities (Meier et al., 2010; Alwmark et al., 2012) and in recent micrometeorites (Olinger et al., 1990, Osawa and Nagao, 2002). Ten grains have Ne ratios that plot more than 2σ to the right of the solar wind fractionation line, and six of these (Ly3-Cr32, Ly3-Cr38, Ly3-Cr41, Ly3-Cr57, Ly4-Cr02 and Ly4-Cr27) even plot more than 3σ to the right of this line. The amount of cosmogenic Ne in these grains is thus clearly resolvable. The majority of the Lynna River SEC grains are rich in trapped Ne; however, the grains with low $^4$He concentrations typically also have a low $^{20}$Ne concentration of around $10^{-8}$ cm$^3$ STP/g. These gas-poor grains do not plot close to the solar wind fractionation line, except perhaps for grain Ly3-Cr47, where both a solar or atmospheric origin are possible. Five gas-poor grains have $^{21}$Ne/$^{22}$Ne ratios of ~0.1 - 0.2: Two of these (Ly3-Cr26 and Ly4-Cr27) plot within uncertainties on the line connecting the solar wind end-members with the cosmogenic Ne point (Figure 3), indicating that they contain traces of solar Ne. Three other grains (Ly3-Cr78, Ly4-Cr3 and Ly4-26), however, contain trapped Ne with a low $^{20}$Ne/$^{22}$Ne ratio of ~6-8, perhaps indicative of the presence of an exotic Ne component, as discussed later. For yet six other gas-poor grains (two from Ly3, four from Ly4; not plotted in Figure 3) the trapped Ne is ill-defined, as the $^{20}$Ne/$^{22}$Ne and $^{21}$Ne/$^{22}$Ne ratios are compatible with zero within 1σ uncertainties. The fraction of grains with low concentrations of trapped gases (12 out of 48) at Lynna River is significantly higher than reported for other localities (e.g., Meier et al., 2010). This is probably related to our preferential selection of large grains (>3 μg; >100 μm diameter), which are more likely to be poor in trapped gases (Meier et al., 2010; Alwmark et al., 2012; this work).

## 3.2. Cosmogenic Ne and cosmic-ray exposure ages

Helium-3 in most grains is dominated by the solar wind contribution. In grains poor in trapped He, $^3$He was either below the detection limit, or the contribution of cosmogenic $^3$He cannot be reliably determined due to the unknown contribution of radiogenic $^4$He. We therefore restrict our discussion to cosmogenic $^{21}$Ne, the contribution of which is calculated with a three-component deconvolution, using the isotopic compositions of solar wind ($^{20}$Ne/$^{22}$Ne = 13.8; $^{21}$Ne/$^{22}$Ne = 0.0329; Heber et al., 2009), fractionated solar wind ($^{20}$Ne/$^{22}$Ne = 11.2, $^{21}$Ne/$^{22}$Ne = 0.0298; Wieler, 2002) and cosmogenic Ne ($^{20}$Ne/$^{22}$Ne, $^{21}$Ne/$^{22}$Ne = ~0.9) as end-members. The resulting concentrations of cosmogenic $^{21}$Ne ($^{21}$Ne$_{cos}$) are listed in Table 1. To calculate cosmic-ray exposure (CRE) ages from these concentrations, we need to consider two different cases: firstly, micrometeoroids that are exposed to cosmic-rays only during transfer from their parent body to Earth, and secondly, micrometeoroids that acquired a large or even dominant part of their cosmogenic inventory during pre-exposure in the regolith of their parent asteroid. Micrometeoroids in transfer to Earth are exposed not only to galactic-cosmic-rays (GCR), but also to solar-cosmic-rays (SCR). The latter can usually be neglected for larger meteoroids because they penetrate only into the uppermost very few cm, which are usually lost upon atmospheric entry. We calculate the production rate of $^{21}$Ne by SCR and GCR with a physical model by Trappitsch and Leya (2013) specifically developed for micrometeoroids, with the target chemistry adjusted for (L chondritic) chromite composition, a spectral rigidity of $R_0$ = 100 MV for SCR, $R_0$ = 550 MV for GCR, and a SCR proton flux ($J_0$) at 1 AU (astronomical unit) of 70 cm$^{-2}$ s$^{-1}$. This flux, which corresponds to the modern value, is stable over the timescales relevant here (Nishiizumi et al., 2009). This model yields a SCR production rate of 12.4 × 10$^{-10}$ cm$^3$ STP $^{21}$Ne/gMa. Since the SCR flux depends on heliocentric distance (with an exponent most likely be-tween -2 and -3; Nishiizumi et al., 1991, and references therein), this production rate needs to be scaled to the typical heliocentric distance at which the micrometeoroid was irradiated. As we will show below, it is unlikely that the micrometeoroids were delivered by Poynting–Robertson drag, given their size and position within the stratigraphic column. Instead, they must have been delivered by the same mechanism as the fossil meteorites found in coeval sediment beds in Sweden, i.e. by direct injection into orbital resonances. The time-averaged heliocentric distance (d = a × (1 + e$^2$/2), where a is the semi-major axis of the orbit and e its eccentricity) of a typical meteoroid orbit reaching from the Earth's orbit to the asteroid belt (e.g., Jenniskens et al., 1992) is between 2 and 3 AU. Taking these factors into account, the scaling factor is likely to be between $3^{-3}$ and $2^{-2}$, i.e., between 0.037 and 0.25, and the SCR production rate is thus (0.5–3.1) × $10^{-10}$ cm$^3$ STP $^{21}$Ne/gMa. From the micrometeoroid model by Trappitsch and Leya (2013), we derive a GCR production rate of 4.5 × $10^{-10}$ cm$^3$ STP $^{21}$Ne/gMa. The combined GCR + SCR production rate for $^{21}$Ne is thus (6.3 ± 1.3) × $10^{-10}$ cm$^3$ STP $^{21}$Ne/gMa. This GCR + SCR production rate is very similar to the empirical $^{21}$Ne production rate of 7.0 × $10^{-10}$ cm$^3$ STP $^{21}$Ne/gMa determined for chromite in decimeter-sized meteorites (Heck et al., 2004; 2008). The lowering of the GCR production rate resulting from the smaller size (and thus negligible production by secondary particles) is just about compensated by additional production from SCR. Since our grains have typical diameters of at least several ten micrometers, recoil losses should be low (<5%; Trappitsch and Leya, 2010) and are therefore neglected here.

Grains Ly3-Cr26 and Ly4-Cr27 show clear contributions of cosmogenic $^{21}$Ne ($^{21}$Ne/$^{22}$Ne-"excesses" relative to the solar-wind fractionation line of 2.1σ and 2.5σ, respectively), and have masses with low relative uncertainties (11.1 ± 0.1 µg and 7.6 ± 0.3 µg, respectively), resulting in well-defined cosmogenic concentrations of 1.6 ± 0.2 × $10^{-10}$ cm$^3$ STP $^{21}$Ne/g and 2.5 ± 0.2 × $10^{-10}$ cm$^3$ STP $^{21}$Ne/g, respectively. Using the combined GCR + SCR production rate from above, we calculate CRE ages of 0.25 ± 0.06 Ma and 0.36 ± 0.08 Ma for grain Ly3-Cr26 and Ly4-Cr27, respectively (these uncertainties do not include the uncertainty of the production rate). These ages are similar to the ones measured in fossil meteorites from coeval sediment beds in southern Sweden (Heck et al., 2004; Heck et al., 2008; Figure 1). This is the first time that such low CRE ages have been resolved in fossil micrometeorites (see discussion section). In addition to these two grains, three grains devoid of solar wind Ne (Ly3-Cr78, Ly4-Cr03 and Ly4-Cr26) also show an isotopic composition indicative of the presence of cosmogenic $^{21}$Ne. Their Ne ratios plot on the same mixing line between cosmogenic Ne and an exotic Ne component (named "Ne-HL + E" in Figure 3) recently observed in chromite grains from recent meteorites (Meier et al., 2014; see discussion section). Since these three grains do not contain any solar wind, it is possible that they derive from decomposed, centimeter-sized fossil meteorites. However, no such grains have so far been found among SEC grains from Swedish and Chinese localities (Meier et al., 2010; Alwmark et al., 2012). More likely, these grains derive from micrometeorites that were just large enough (before entering the Earth's atmosphere) to completely enclose the chromite grains in their interior, thus shielding them from solar wind exposure, while still being small enough to escape melting upon atmospheric entry (Love and Brownlee, 1991). Using the GCR + SCR production rate from above, CRE ages of 0.14 ± 0.06 Ma, 0.51 ± 0.26 Ma and 0.34 ± 0.08 Ma result for grains Ly3-Cr78, Ly4-Cr03 and Ly4-Cr26, respectively.

The second class of grains with clear cosmogenic $^{21}$Ne excesses are rich in solar wind-derived Ne. Their significantly higher concentrations of $^{21}$Ne$_{cos}$ correspond to CRE ages of up to several ten Ma (regardless of whether 2π or 4π production rates are used; see below). Since we know, from their abundance and position in the sediment column, that they must have been released in the LCPB break-up event, these grains must have acquired most of their cosmogenic Ne prior to their journey to Earth as micrometeoroids, which is only possible if they resided in the regolith layer of the LCPB asteroid (cf., Meier et al., 2010; discussion below). To calculate regolith residence times, we first calculate the value $^{21}$Ne$_{exc}$ = $^{21}$Ne$_{cos}$ - $^{21}$Ne$_{cos(Ly3-Cr26)}$ for grains from Ly3, and $^{21}$Ne$_{exc}$ = $^{21}$Ne$_{cos}$ -

$^{21}$Ne$_{cos(Ly4-Cr27)}$ for grains from Ly4. We then use the 2π-irradiation model by Leya et al. (2001) to calculate the maximum $^{21}$Ne production rate of 3.2 × 10$^{-10}$ cm$^3$ STP $^{21}$Ne/gMa under regolith exposure, corresponding to a shielding depth of about 40–45 g/cm$^2$. The resulting minimum regolith CRE ages, of up to 50 Ma, are given in Table 1. True regolith exposure ages may be considerably higher than these minimum values, depending on the actual average burial depth of the grain over its exposure history in the regolith.

## 4. DISCUSSION

The general properties of He and Ne in SEC grains from mid-Ordovician sediments have been discussed before, e.g. in Meier et al. (2010) and Alwmark et al. (2012). In the following, we will therefore focus on the novel findings, namely the determination of well-defined, short (<1 Ma) CRE ages in five individual SEC grains, and the stratigraphic implications of that finding. We will also briefly discuss the grains with well-resolved, significantly longer (1–50 Ma) CRE ages.

### 4.1. Grains with short cosmic-ray exposure ages

*4.1.1. Overview*
The five grains with well-defined CRE ages shorter than 1 Ma are listed in Table 2. The two grains from bed Ly3 have CRE ages of 0.14 ± 0.06 Ma and 0.25 ± 0.06 Ma, respectively, and three grains from bed Ly4 have CRE ages of 0.36 ± 0.08 Ma, 0.51 ± 0.26 Ma and 0.34 ± 0.08 Ma, respectively. The CRE ages of the two grains from Ly3 are marginally inconsistent with each other within their 1σ uncertainty, while the CRE ages of the three grains from Ly4 agree with each other. As observed before in chromite grains from fossil meteorites from southern Sweden (Heck et al., 2004; 2008; Figure 1), SEC grains from the stratigraphically younger sediment bed Ly4 have a longer mean CRE age (0.36 ± 0.06 Ma) than the grains from the stratigraphically older bed Ly3 (0.19 ± 0.04 Ma). As was concluded for the fossil meteorites, this observation is consistent with the idea that all grains started their exposure to cosmic-rays at the same point in time, but were delivered to Earth after journeys of different duration. As the stratigraphic position, the elemental composition and abundance of the grains show (Korochantsev et al., 2009; Lindskog et al., 2012), the starting point in time can be identified with the break-up of the L chondrite parent body (LCPB) asteroid. This is the first time that such low and well-resolved CRE ages have been determined in SEC grains/fossil micrometeorites. The CRE ages are still lower than the lowest (resolved) $^{21}$Ne GCR + SCR exposure ages determined for recent micrometeorites (~0.4–1.0 Ma; Olinger et al., 1990), and similar to the shortest $^{10}$Be/$^{26}$Al-based GCR + SCR exposure ages measured in antarctic micrometeorites by Nishiizumi et al. (1991). Since SEC grains are much more abundant, and thus easier to find, than fossil meteorites, this finding opens up the possibility of precisely correlating, on a worldwide basis, beds deposited in the few million years after the break-up of the L chondrite parent body asteroid.
Four out of five SEC grains with short CRE ages have a high mass (>4 μg). The ratio of the solar wind Ne to cosmogenic Ne usually should scale with the inverse radius of a grain, making large grains somewhat more favorable for the detection of cosmogenic Ne. However, the five grains with short CRE ages either have significantly less solar-wind-derived Ne than would be expected from such scaling (Ly3-Cr26 and Ly4-Cr27), or no solar wind Ne at all (Ly3-Cr78, Ly4-Cr03 and Ly4-Cr26). The first two of these grains appear thus to have lost a relatively large portion of their solar wind gases because they were more strongly heated in the atmosphere than smaller grains (Love and Brownlee, 1991). The grains with no solar wind might have lost all of their solar wind gases upon atmospheric entry, or were shielded completely from solar wind exposure by surrounding material. For future work, when attempting to extend the measurement of short CRE ages in SEC grains both geographically and in time, high mass (>4 μg) SEC grains appear to offer the best prospects: they tend to be gas-poor, and the larger mass makes it more likely that a short CRE age can be resolved.

*4.1.2 Reconstruction of sedimentation rates*
The determination of sedimentation rates (S) from fossil (micro)meteorites requires two inputs: a difference in stratigraphic height (Δh) and a difference in deposition time (Δt), in the manner:

$$S = \Delta h / \Delta t$$

Since the SEC grains have been recovered from the bulk samples, it is difficult to assign them a clear stratigraphic height. The only information we have is the thickness of the beds from which they were sampled (~15 cm and ~5 cm for Ly3 and Ly4, respectively). If the grains were deposited in the beds within which they were found (i.e., they were not re-deposited or deposited into lower-lying beds by bioturbation), their stratigraphic height difference is Δh = 10 ± 10 cm. The difference in deposition time (Δt) is then the difference in the average CRE ages of grains found in the two beds, 0.16 ± 0.07 Ma, resulting in a sedimentation rate of 0.61 ± 0.66 m/Ma. The sedimentation rate is consistent with zero only because the two sediment beds are directly adjacent. Even at the upper limit (Δh = 20 cm, Δt = 0.1 Ma), the resulting sedimentation rate at Lynna River (2 m/Ma) is still a factor of ~2 lower than the value determined for the Thorsberg quarry in Sweden using chromite from fossil meteorites (3.8 m/Ma). This observation is compatible with the significantly higher concentration of SEC grains at Lynna River, compared to Swedish localities.

*4.1.3. Delivery of SEC grains from the asteroid belt to Earth*
The delivery of centimeter- to meter-sized objects from the asteroid belt is separated into two phases (Bottke et al., 2002). First, slow migration, driven by the Yarkovsky effect, from the parent body asteroid to a near-by orbital resonance with Jupiter (and, for the ν6 resonance, Saturn), lasting about 1–100 Ma. After entrance into an orbital resonance, the meteoroid orbit is excited to high eccentricities, until it (most likely) collides with the sun or one of the terrestrial planets. This second phase lasts only 0.1–1 Ma. The very low CRE exposure ages (0.1–1 Ma) of fossil meteorites found in mid-Ordovician sediments indicate that they were injected directly into a resonance. This is possible if they were formed in a very large collision that produced copious amounts of debris, so that a sizable fraction was likely to start its journey to Earth on a resonant orbit (Gladman et al., 1997; Heck et al., 2004).

In contrast to larger meteoroids, the transport of micrometeoroid-sized (~10 μm to ~1 cm) objects from their source regions to Earth is usually mainly achieved by Poynting–Robertson (PR) drag (e.g., Burns et al., 1979). Using the empirical formulas given by Burns et al. (1979), and putting the starting heliocentric distance to 2.8 AU, the typical semi-major axis of asteroids belonging to the Gefion family, which has likely been produced by the LCPB break-up event (Nesvorný et al., 2009), we calculate PR transfer times for each of the SEC grains in Table 2. These transfer times are minimum values, as the original micrometeorites might have been larger (i.e., they might not have consisted exclusively of chromite, but of an assemblage of additional minerals, including less weathering-resistant types that were not recovered with the SEC grains), resulting in larger transfer times. These minimum PR transfer times are always larger than the observed CRE ages, indicating that these SEC grains were not mainly delivered by PR drag. Instead, they must have been injected directly into an orbital resonance, like the fossil meteorites. This means that it should be possible to measure short CRE ages in SEC grains both close to the break-up event (resonance-delivered grains) and up to several Ma later (PR drag-delivered grains).

*4.1.4. Implications for local stratigraphy*
A discussion of the full implications of our result for the local stratigraphy at Lynna River is beyond the scope of this paper. Nevertheless, we note that the CRE ages of fossil micrometeorites clearly link some beds at the Lynna River section to coeval beds in the fossil-meteorite-rich strata at the Thorsberg quarry section in Sweden. The bed-to-bed correlation depends on whether we directly compare CRE ages of fossil meteorites and SEC grains, in which case Ly3 is stratigraphically linked to the "Arkeologen" unit, while Ly4 is linked to the "Golvsten" unit (see Figure 1). Alternatively, we can base the correlation on the inferred average deposition rate for the whole section (3.8 ± 0.8 m/Ma), in which case the counterpart of Ly3 is the "Golvsten" bed, and Ly4 is likely to be located somewhere within the lower part of the gray "Täljsten" interval (see Eriksson et al., 2012; Figure 1).

*4.1.5. Exotic Ne?*
The three gas-poor grains Ly3-Cr78, Ly4-Cr03 and Ly4-Cr26 not only have well-defined cosmogenic $^{21}$Ne concentrations, but also plot below the mixing lines between the cosmogenic point and both solar or atmospheric end-members in a Ne three-isotope diagram (Figure 3). Given the large uncertainties, this would perhaps not be too conspicuous if a similar observation of an unknown or "exotic" Ne-component in ordinary chondritic chromite from recent meteorites had not recently been reported (Meier et al., 2014). The Ne isotopic compositions of chromite grains from Ghubara (L5; Meier et al., 2014), Mt. Tazerzait (L5), Harleton (L6), Saint-Séverin (LL6), Eva (H5 and Hessle (H5; Heck et al., 2008) all plot roughly on a line connecting the cosmogenic point with an end-member tentatively named "HL + E" by Meier et al. (2014) for its position in between the exotic (presolar) components Ne-HL and Ne-E. The typical contribution of this hypothetical component, on the order of ~$10^{-14}$ cm$^3$ STP $^{22}$Ne per grain, is similar to the value determined for the three SEC grains, 0.4 - 0.8 × $10^{-14}$ cm$^3$ STP $^{22}$Ne per grain. The three SEC grains, with a lower CRE age and thus a significantly lower contribution of cosmogenic $^{21}$Ne compared to the meteorite grains, extend the mixing line further towards the "HL + E" end-member, thus providing further support for this finding.

**4.2. Grains pre-exposed in the regolith**

For nine out of 14 SEC grains with well-defined (>2σ) concentrations of cosmogenic $^{21}$Ne, this concentration significantly exceeds the value they could have acquired during their journey towards Earth, sometimes by up to two orders of magnitude. These grains must have acquired the bulk of their cosmogenic Ne not in transit to Earth, but while residing in the regolith of the pre-break-up LCPB asteroid. In Figure 4, we plot the $^{21}$Ne$_{exc}$ concentration of each of these 14 grains versus its concentration of (almost exclusively solar) $^{20}$Ne. Grains rich in solar Ne tend to have higher $^{21}$Ne$_{cos}$ values, as observed in SEC grains from other localities before (Meier et al., 2010; Alwmark et al., 2012). A similar correlation between these two parameters has also been observed in regolith breccia meteorites (e.g., in Fayetteville (H4); Wieler et al., 1989), and has been interpreted as a signature of regolith irradiation; the more time the grains spend in the topmost ~1–2 m of a well-mixed regolith (where they are exposed to galactic cosmic-rays), the more frequently they will also be exposed to solar wind ions at the immediate surface of the regolith. In Figure 4, as a comparison, we use values measured in bulk samples of Fayetteville (Wieler et al., 1989), subtract the $^{21}$Ne$_{cos}$ produced during the 4π irradiation of that meteorite (i.e., during its transfer to Earth; 7.0 × $10^{-8}$ cm$^3$ STP/g) and scale the remaining $^{21}$Ne$_{cos}$ down by a factor of ~5 to account for the different $^{21}$Ne production rates in a bulk ordinary chondrite vs. chromite (thereby predicting the $^{21}$Ne$_{cos}$ produced dur-ing 2π irradiation in chromite grains from Fayetteville). The slope of the correlation line of the "regolith trend" in Fayetteville is steeper than the one observed for SEC grains (Figure 4), i.e., for a given time the grains spend in the topmost ~1–2 m of the regolith, the SEC grains acquire about five times higher concentrations of solar wind gases, compared to the bulk samples from Fayetteville. This might be related to gas-loss experienced by regolith breccias during shock-induced compaction and lithification, and in this case would require a loss of solar $^{20}$Ne on the order of ~80%, if the "regolith-trend" on the H and L chondrite parent body initially had a similar slope.

A fraction of 25% regolith-derived SEC grains might apear very high, given that only the topmost few meters (a volumetrically insignificant fraction of the LCPB asteroid) are directly exposed to GCR. However, this ratio only needs to reflect the ratio of material ejected from the former regolith layer versus new, interior-derived dust produced directly during the collision. The bulk of the interior (non-regolith) material is still contained in the asteroids of the Gefion family formed by the break-up of the LCPB (Nesvorný et al., 2009). Furthermore, the LCPB has probably been disrupted and reassembled in the early solar system (Taylor et al., 1987). The LCPB interior might thus also have contained grains from an early regolith that were incorporated during re-assembly, and re-released during the break-up 470 Ma ago.

## 5. CONCLUSIONS AND OUTLOOK

We have analyzed the He and Ne concentration of 50 sediment-dispersed chromite grains of extraterrestrial (L chondritic) major element composition (SEC grains) recovered from ca. 470 Ma old sedimentary rocks from the Lynna River section in Russia. These SEC grains are shown to be fossil micrometeorites, i.e., the surviving part of micrometeoroid-sized dust produced in break-up of the L chondrite parent body (LCPB) asteroid. Most of the grains contain high concentrations of solar wind He and Ne, similar to what has been reported in SEC grains from coeval sediments in Sweden and China by Meier et al. (2010) and Alwmark et al. (2012). We find five relatively massive grains (>4 μg) with very low solar-wind-derived Ne concentrations and a well-resolved excess of cosmogenic $^{21}$Ne ($^{21}$Ne/$^{22}$Ne = 0.1–0.2). Using galactic and solar cosmic-ray production rates from a recent model (Trappitsch and Leya, 2013), we calculate cosmic-ray exposure (CRE) ages for these five micrometeorites, of 0.14 to ~0.51 Ma. These CRE ages are shorter than the expected Poynting–Robertson transfer times for grains of this size, suggesting that they, like the fossil meteorites, were injected directly into an orbital resonance. This is the first time that such short, well-resolved $^{21}$Ne-based CRE ages (<1 Ma) have been measured in (fossil or recent) micrometeorites. These CRE ages are similar to those measured by Heck et al. (2004) in fossil meteorites found in coeval sediments in Sweden, thus allowing the determination of sedimentation rates at the Lynna River section, and the cross-correlation of coeval sediment beds independent of lateral distance. Three of the five SEC grains with short CRE ages have a Ne isotopic composition consistent with the presence of an exotic, $^{22}$Ne-rich component identified in chromite grains from recent meteorites (Meier et al., 2014), giving further support to that discovery. We furthermore find nine grains with well-resolved cosmic-ray exposure ages of up to 50 Ma, which is the time these grains resided in the regolith of the LCPB, prior to its break-up. A clear "regolith trend" of correlated cosmogenic $^{21}$Ne and solar $^{20}$Ne is observed for these grains, as seen before in regolith-breccia meteorites (e.g., Fayetteville).

## ACKNOWLEDGMENTS

This work was supported by the Swiss National Science Foundation (grant to MM) and the Swedish Research Council (grant to BS). Andrei Dronov is thanked for invaluable guidance during the sampling fieldwork at Lynna River, and Reto Trappitsch is thanked for providing, and helping with the micrometeorite production rate models. We thank the associate editor Gregory Herzog, as well as reviewers David Nesvorný, Tim Swindle and an anonymous third reviewer for their careful and helpful reviews. *Author contributions:* MM and BS developed the project. MM did the noble gas analysis and wrote the main text. AL prepared the SEC grains from their sediment beds, and measured their elemental composition. CM and RW contributed their expertise in mass spectrometry and noble gas cosmochemistry, respectively. All authors discussed the results and text at all stages of development.

# FIGURES

**Figure 1: Fossil meteorite exposure ages in relation to stratigraphic position**

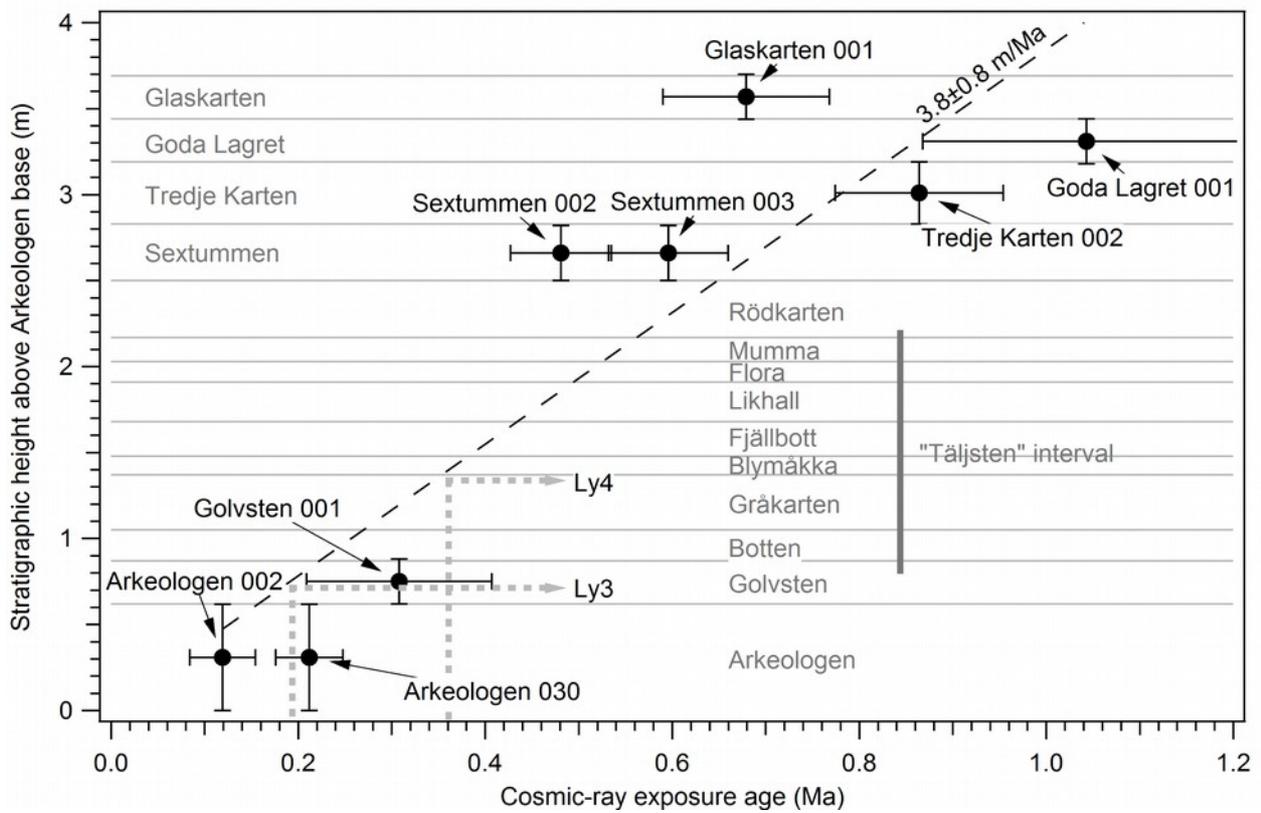

**Figure 2:** $^3$He/$^4$He vs. $^4$He concentration

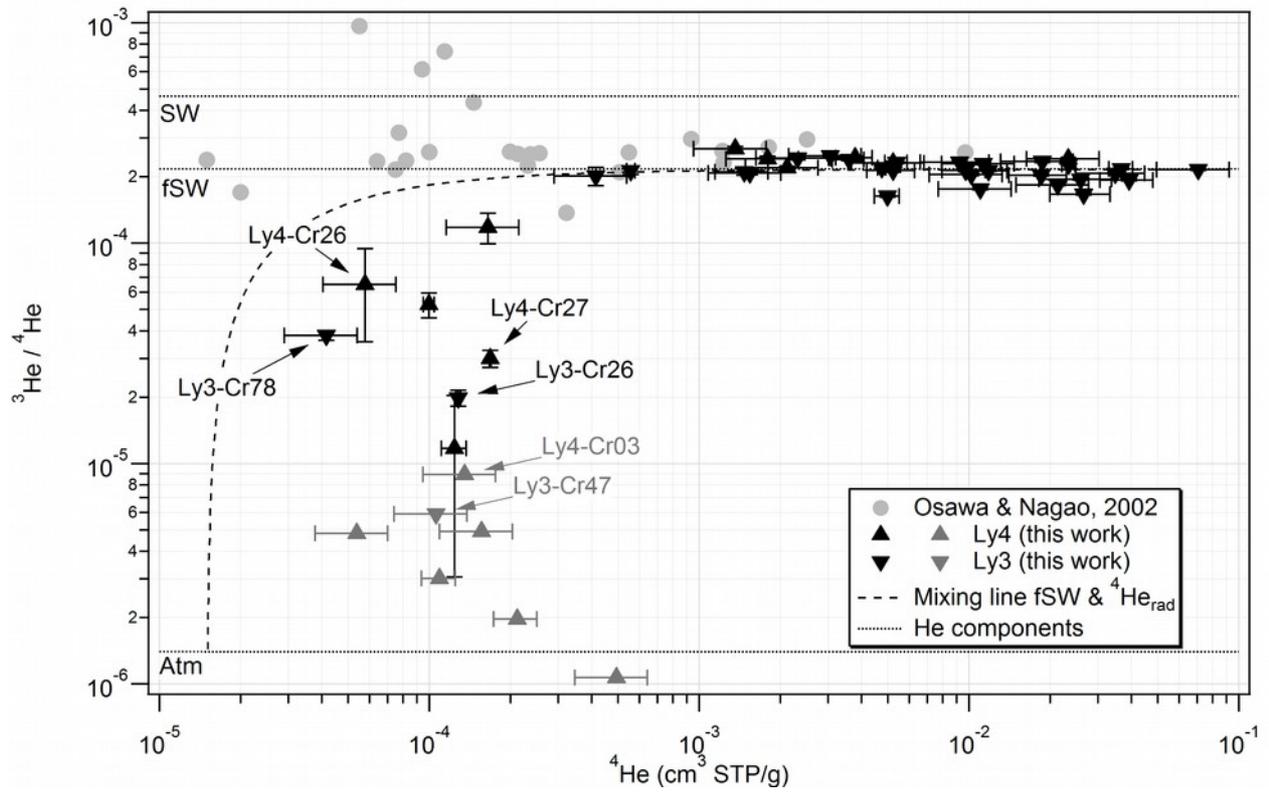

**Figure 3: Ne three-isotope diagram**

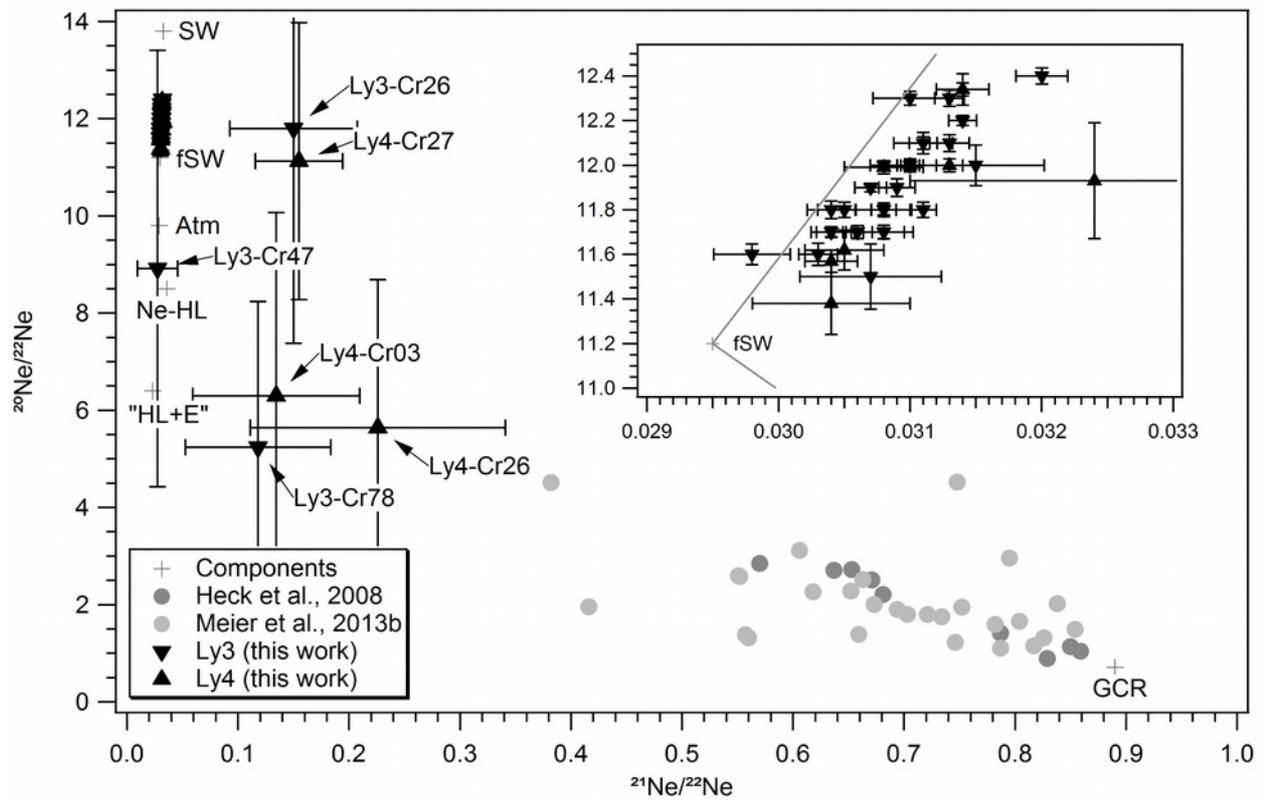

**Figure 4: Regolith pre-exposure signature**

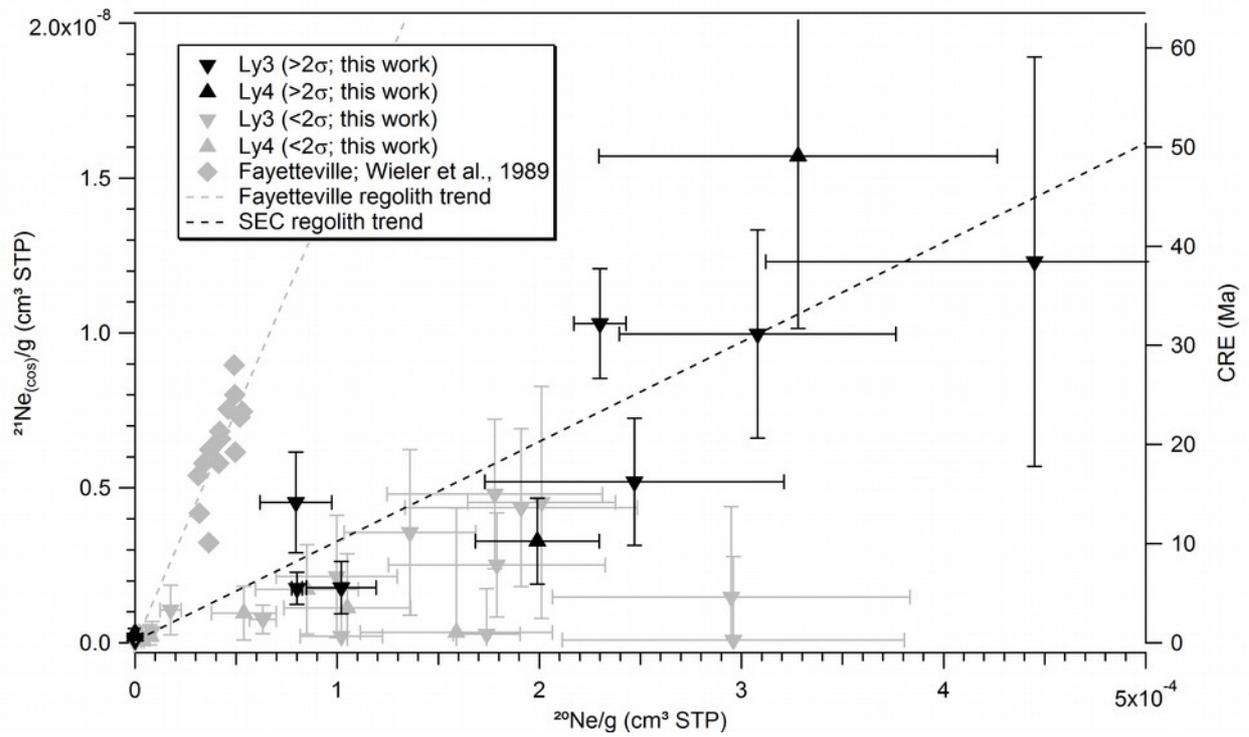

# FIGURE CAPTIONS

**Figure 1: Fossil meteorite exposure ages in relation to stratigraphic position**

Cosmic-ray exposure ages and stratigraphic positions of 8 different fossil meteorites (represented by black squares) from Heck et al. (2004). The sediment beds (units) are labeled according to their traditional names given by the quarry-workers. The long-dashed line indicates one possible sedimentation scenario, roughly compatible with the measured cosmic-ray exposure ages of the fossil meteorites: a constant sedimentation rate of about 3.8 m/Ma (or 3.8 mm/ka). Note that for fossil meteorites, the concentration of cosmogenic $^{21}$Ne also depends on shielding conditions – burial depth, size of the meteoroid – during their transfer to Earth, inducing additional variability. The "Täljsten" interval (a sequence of gray limestone within otherwise red-colored limestone) is indicated. From the CRE ages of the Ly3 and Ly4 grains, these two beds can now be cross-correlated (as shown by the dashed arrows).

**Figure 2: $^3$He/$^4$He vs. $^4$He concentration**

The isotopic composition of He plotted against the $^4$He concentration, for both Lynna River grains and recent micrometeorites from Antarctica (Osawa and Nagao, 2002). Most of the grains plot close to the fractionated solar wind (fSW; formerly "solar energetic particles" = SEP) value in $^3$He/$^4$He. For Lynna River grains represented by gray triangles, only $^4$He was detected, and the $^3$He/$^4$He ratio indicated is an upper limit. The black, dotted curve represents a mixing line between fSW and radiogenic $^4$He with a typical ordinary chondritic concentration of $1.5 \times 10^{-5}$ cm$^3$ STP/g. The six labeled grains are the ones labeled in Figure 3 as well, for comparison.

**Figure 3: Ne three-isotope diagram**

Measured ratios $^{20}$Ne/$^{22}$Ne vs. $^{21}$Ne/$^{22}$Ne for all analyzed Lynna River grains. Most of the grains plot close to a line connecting the solar wind (SW) with the fractionated solar wind (fSW) component (shown in inset), indicating the Ne in these grains derives from implantation by the solar wind. Most grains plot slightly to the right of this fractionation line, indicating the presence of some cosmogenic Ne (cos). Grains Ly3-Cr26 and Ly4-Cr27 both show a clear excess of cosmogenic $^{21}$Ne. Some of the gas-poor grains with well-defined errors plot below the line connecting the solar-wind and cosmogenic end-members, perhaps indicating the presence of a $^{22}$Ne-rich component (here labeled "HL + E") in the grains (see main text), previously identified in recent ordinary chondrites analyzed by Heck et al. (2008) and Meier et al. (2013b). The data points from these studies are shown as gray dots.

**Figure 4: Regolith pre-exposure signature**

Cosmogenic $^{21}$Ne in Lynna River SEC grains vs. $^{20}$Ne concentration. Fourteen grains have well-defined (>2σ) excesses of cosmogenic $^{21}$Ne in the Ne-three-isotope diagram (black triangles; five of the grains plot very close to the origin). Grains Ly3-26 and Ly4-27, which have well-defined CRE ages of 0.25±0.06 and 0.36±0.08 Ma, respectively, plot near the origin in this diagram. Grains with large uncertainties in cosmogenic $^{21}$Ne contributions are given with gray triangle symbols. From the 14 grains with well-defined cosmogenic contributions, a regression line ($R^2$ = 0.82; black, dashed) can be calculated and compared with similar "regolith trends" in regolith breccia meteorites, e.g., Fayetteville (H4; Wieler et al., 1989), the data of which are shown here as gray diamonds. The Fayettville data have been scaled to account for the different production rates of cosmogenic $^{21}$Ne in bulk ordinary chondrites and chromite.

# TABLES

## Table 1: Mass, elemental and He, Ne composition

| Grain | Mass (μg) | Mg wt% | Al | Ti | V | Cr | Fe | Mn | Zn | Total | $^3$He/$^4$He (× 10$^4$) | $^4$He/g (× 10$^5$) | $^{20}$Ne/$^{22}$Ne | $^{21}$Ne/$^{22}$Ne | $^{20}$Ne/g (× 10$^5$) | $^4$He/$^{20}$Ne | $^{21}$Ne$_{cos}$ (× 10$^8$) | CRE, 2π (Ma) |
|---|---|---|---|---|---|---|---|---|---|---|---|---|---|---|---|---|---|---|
| **Lynna River, Ly3** | | | | | | | | | | | | | | | | | | |
| Ly3-Cr01 | 0.7±0.2 | 2.6 | 6.7 | 1.8 | 0.7 | 55.0 | 28.1 | n.d. | 1.1 | 96.0 | 2.33±0.02 | 92$_0$±26$_0$ | 11.56±0.05 | 0.0298±0.0003 | 12±3 | 79.1±0.1 | <0 | <0 |
| Ly3-Cr02 | 1.9±0.6 | 2.1 | 5.1 | 2.6 | 0.8 | 43.0 | 43.2 | n.d. | n.d. | 96.8 | 1.76±0.01 | 11$_{00}$±3$_{00}$ | * | 0.0313±0.0009 | * | * | n.d. | n.d. |
| Ly3-Cr03 | 1.4±0.4 | 1.3 | 7.6 | 3.0 | 1.3 | 61.3 | 20.0 | n.d. | 2.2 | 96.7 | (grain lost during lasering) | | | | | | | |
| Ly3-Cr04 | 1.3±0.3 | 1.7 | 6.9 | 2.5 | 1.4 | 59.5 | 23.1 | 1.3 | 0.3 | 96.7 | 2.03±0.01 | 18$_{00}$±4$_{00}$ | 11.73±0.03 | 0.0308±0.0002 | 14±3 | 136.4±0.2 | 0.36±0.27 | 10.6 |
| Ly3-Cr05 | 0.4±0.1 | 1.7 | 6.2 | 1.3 | 0.8 | 58.9 | 16.9 | n.d. | 14.1 | 99.9 | $^3$He n.d. | 17±5 | 11±28 | 0.12±0.31 | 5.4±2.9$^a$ | 31$_{00}$±14$_{00}$ | n.d. | n.d. |
| Ly3-Cr07 | 0.8±0.2 | 2.0 | 6.1 | 1.8 | 0.8 | 51.1 | 35.4 | n.d. | 0.5 | 97.7 | 2.29±0.03 | 11$_{00}$±3$_{00}$ | 11.86±0.04 | 0.0309±0.0001 | 19±6 | 59.1±0.2 | 0.44±0.25 | 13.1 |
| Ly3-Cr10 | 2.1±0.2 | 1.4 | 7.9 | 3.7 | 1.2 | 58.1 | 26.9 | n.d. | n.d. | 99.2 | 2.32±0.02 | 54$_0$±5$_0$ | 12±0.01 | 0.0308±0.0001 | 17±2 | 30.9±0.0 | 0.03±0.15 | 0.4 |
| Ly3-Cr11 | 0.8±0.3 | 3.7 | 14.2 | 0.9 | 0.9 | 55.3 | 25.9 | n.d. | n.d. | 100.9 | 1.95±0.02 | 26$_{00}$±8$_{00}$ | 11.74±0.02 | 0.0304±0.0002 | 36±11 | 71.6±0.1 | <0 | <0 |
| Ly3-Cr12 | 1.1±0.3 | 1.7 | 7.0 | 2.4 | 1.0 | 59.5 | 27.1 | n.d. | n.d. | 98.7 | 2.05±0.02 | 10$_{00}$±3$_{00}$ | 11.85±0.02 | 0.0307±0.0001 | 29±9 | 34.8±0.0 | 0.15±0.29 | 4.1 |
| Ly3-Cr15 | 0.5±0.1 | 1.1 | 5.3 | 2.2 | 0.8 | 54.0 | 33.0 | n.d. | 0.7 | 97.1 | 2.17±0.02 | 37$_{00}$±4$_{00}$ | 11.81±0.04 | 0.0304±0.0002 | 40±4 | 91.2±0.2 | <0 | <0 |
| Ly3-Cr16 | 1.9±0.6 | 2.0 | 6.1 | 1.9 | 0.9 | 50.9 | 36.3 | 1.1 | 0.3 | 99.5 | 2.49±0.04 | 31$_0$±9$_0$ | 12.08±0.03 | 0.0311±0.0002 | 10±3 | 30.7±0.1 | 0.21±0.2 | 6.3 |
| Ly3-Cr21 | 0.8±0.2 | 2.1 | 6.6 | 2.0 | 0.8 | 48.5 | 38.9 | n.d. | 0.4 | 99.3 | 1.67±0.02 | 27$_{00}$±7$_{00}$ | 12.29±0.03 | 0.0310±0.0003 | 29±7 | 90.5±0.2 | <0 | <0 |
| Ly3-Cr23 | 3±0.5 | 2.4 | 7.3 | 3.6 | 1.3 | 62.1 | 21.6 | n.d. | 1.8 | 100.1 | 2.13±0.02 | 12$_{00}$±2$_{00}$ | 12.08±0.05 | 0.0311±0.0001 | 10±2 | 115.3±0.4 | 0.18±0.08 | 5.0 |
| Ly3-Cr24 | 0.5±0.1 | 1.9 | 8.3 | 3.1 | 0.8 | 62.1 | 18.4 | n.d. | 2.3 | 96.9 | 2.16±0.02 | 71$_{00}$±21$_{00}$ | 11.72±0.03 | 0.0308±0.0002 | 44±13 | 158.9±0.1 | 1.2±0.7 | 37.5 |
| Ly3-Cr26 | 11.1±0.2 | 1.8 | 5.9 | 3.0 | 0.9 | 62.6 | 21.4 | n.d. | 3.0 | 98.6 | 0.199±0.017 | 12.8±0.2 | 11.81±4.42 | 0.15±0.06 | 1.5±0.1$^a$ | 825$_0$±63$_0$ | 0.016±0.002 | ≡0 |
| Ly3-Cr27 | 0.9±0.3 | 2.7 | 7.1 | 3.2 | 1.1 | 61.3 | 22.9 | n.d. | 1.4 | 99.7 | 2.35±0.02 | 19$_{00}$±6$_{00}$ | 12.15±0.04 | 0.0313±0.0001 | 18±5 | 105.3±0.2 | 0.48±0.24 | 14.4 |
| Ly3-Cr28 | 1.4±0.1 | 2.8 | 5.3 | 3.1 | 0.6 | 57.1 | 27.9 | n.d. | n.d. | 96.8 | 2.13±0.08 | 56±2 | 11.51±0.15 | 0.0307±0.0005 | 0.83±0.03 | 67.3±0.2 | 0.037±0.032 | 0.7 |
| Ly3-Cr29 | 1.9±0.2 | 2.5 | 6.9 | 2.8 | 0.8 | 59.6 | 25.0 | n.d. | 0.7 | 98.3 | 1.63±0.02 | 50$_0$±5$_0$ | 12.29±0.04 | 0.0313±0.0001 | 6.3±0.7 | 79.0±0.1 | 0.076±0.046 | 1.9 |
| Ly3-Cr31 | 0.7±0.2 | 1.5 | 6.1 | 2.6 | 0.9 | 53.0 | 35.2 | n.d. | n.d. | 99.3 | 2.08±0.02 | 35$_{00}$±10$_{00}$ | 11.66±0.02 | 0.0304±0.0001 | 30±8 | 117.0±0.2 | 0.01±0.27 | n.d. |
| Ly3-Cr32 | 1±0.3 | 1.9 | 6.4 | 1.5 | 0.7 | 54.4 | 30.9 | n.d. | 0.5 | 96.3 | 1.84±0.01 | 21$_{00}$±6$_{00}$ | 11.97±0.03 | 0.0310±0.0001 | 25±7 | 86.5±0.2 | 0.52±0.2 | 15.6 |
| Ly3-Cr38 | 0.9±0.2 | 2.4 | 6.6 | 3.8 | 0.9 | 60.8 | 21.3 | n.d. | 1.1 | 96.9 | 1.94±0.01 | 39$_{00}$±9$_{00}$ | 12.17±0.02 | 0.0314±0.0001 | 31±7 | 127.1±0.1 | 1±0.34 | 30.6 |
| Ly3-Cr40 | 1±0.2 | 1.4 | 4.2 | 1.7 | 0.9 | 55.3 | 34.7 | n.d. | n.d. | 98.2 | 2.14±0.04 | 52$_0$±10$_0$ | 11.76±0.03 | 0.0305±0.0002 | 10±2 | 51.6±0.1 | 0.02±0.18 | 0.2 |
| Ly3-Cr41 | 0.9±0.1 | 1.6 | 6.4 | 3.1 | 0.9 | 64.3 | 22.1 | n.d. | 1.5 | 99.9 | 2.27±0.02 | 23$_{00}$±1$_{00}$ | 11.83±0.03 | 0.0311±0.0001 | 23±1 | 101.7±0.2 | 1±0.2 | 31.3 |
| Ly3-Cr43 | 1.1±0.2 | 4.5 | 13.3 | 0.6 | 1.3 | 56.2 | 24.5 | n.d. | n.d. | 100.4 | 2.14±0.03 | 97$_0$±18$_0$ | 11.76±0.02 | 0.0308±0.0002 | 20±4 | 48.4±0.1 | 0.45±0.37 | 13.8 |

| Sample | | | | | | | | | | | | | | | | | |
|---|---|---|---|---|---|---|---|---|---|---|---|---|---|---|---|---|---|
| Ly3-Cr46 | 1.5±0.5 | 3.1 | 7.1 | 2.3 | 0.7 | 47.0 | 40.7 | n.d. | n.d. | 100.9 | 2.43±0.04 | $23_0 \pm 7_0$ | 11.72±0.03 | 0.0306±0.0001 | 18±5 | 13±0 | 0.25±0.17 | 7.2 |
| Ly3-Cr47 | 0.7±0.2 | 5.0 | 6.9 | 1.6 | 1.0 | 56.2 | 27.2 | n.d. | n.d. | 97.9 | $^3$He n.d. | 11±3 | 8.9±4.5 | 0.028±0.018 | 7.8±3.0[a] | $136_0 \pm 32_0$ | 0.000±0.013 | ~0 |
| Ly3-Cr50 | 0.4±0.1 | 3.7 | 7.2 | 3.1 | 0.8 | 59.0 | 23.9 | n.d. | 0.4 | 98.1 | 2.02±0.19 | 41±12 | 11.97±0.09 | 0.0315±0.0005 | 1.7±0.5 | 23.7±0.3 | 0.11±0.08 | 2.8 |
| Ly3-Cr52 | 1.5±0.2 | 1.5 | 6.3 | 2.3 | 0.6 | 57.8 | 31.1 | n.d. | n.d. | 99.6 | 2.39±0.03 | $36_0 \pm 5_0$ | 11.64±0.05 | 0.0303±0.0001 | 6.3±0.8 | 57.4±0.2 | <0 | <0 |
| Ly3-Cr54 | 2.6±0.8 | 3.1 | 8.4 | 2.6 | 0.7 | 44.6 | 38.8 | 1.0 | n.d. | 99.2 | 2.08±0.04 | $15_0 \pm 5_0$ | * | 0.0294±0.0005 | * | * | n.d. | n.d. |
| Ly3-Cr57 | 0.9±0.2 | 5.4 | 4.0 | 1.6 | 0.6 | 53.9 | 35.0 | n.d. | n.d. | 100.5 | 2.11±0.07 | $15_0 \pm 3_0$ | 12.44±0.04 | 0.032±0.0002 | 7.9±1.8 | 18.5±0.1 | 0.45±0.16 | 13.8 |
| Ly3-Cr58 | 1.5±0.1 | 1.3 | 4.5 | 2.3 | 0.6 | 52.8 | 34.5 | n.d. | 0.4 | 96.4 | 2.20±0.03 | $48_0 \pm 2_0$ | 11.82±0.03 | 0.0308±0.0001 | 8±0.3 | 59.4±0.1 | 0.18±0.05 | 5.0 |
| Ly3-Cr60 | 1.1±0.3 | 5.3 | 7.4 | 2.6 | 0.9 | 60.7 | 18.1 | n.d. | 2.8 | 97.8 | $^3$He n.d. | 10±3 | 50±1680 | 3±109 | 0.26±0.56[a] | <120000 | n.d. | n.d. |
| Ly3-Cr78 | 8.9±2.7 | 4.3 | 5.0 | 2.7 | 1.1 | 62.1 | 17.0 | n.d. | 5.5 | 97.7 | 0.38±0.02 | 4.2±1.2 | 5.2±3.0 | 0.12±0.07 | 0.48±0.17[a] | $862_0 \pm 160_0$ | 0.0089±0.0031 | ~0 |
| Ly3-Cr79 | 8.8±0.4 | 6.2 | 7.4 | 2.9 | 0.9 | 60.7 | 15.8 | n.d. | 6.7 | 100.6 | (grain lost during lasering) | | | | | | | |
| **Lynna River, Ly4** | | | | | | | | | | | | | | | | | | |
| Ly4-Cr01 | 1.5±0.5 | 1.3 | 7.7 | 3.7 | 0.9 | 65.1 | 17.8 | n.d. | n.d. | 96.5 | 2.68±0.06 | $14_0 \pm 4_0$ | 12.34±0.07 | 0.0314±0.0002 | 5.4±1.6 | 25.2±0.1 | 0.096±0.087 | 2.3 |
| Ly4-Cr02 | 0.7±0.2 | 1.9 | 6.4 | 2.2 | 0.9 | 44.2 | 40.0 | n.d. | n.d. | 95.6 | 2.41±0.02 | $23_{00} \pm 7_{00}$ | 12±0.03 | 0.0313±0.0001 | 33±10 | 71±0.1 | 1.6±0.6 | 46.9 |
| Ly4-Cr03 | 1.5±0.4 | 2.6 | 6.6 | 2.7 | 1.2 | 63.0 | 24.5 | n.d. | n.d. | 100.6 | $^3$He n.d. | 13±4 | 6.3±3.8 | 0.14±0.08 | 1.8±0.8[a] | $75_{00} \pm 27_{00}$ | 0.032±0.015 | 0.3 |
| Ly4-Cr04 | 1.8±0.5 | 2.3 | 7.9 | 3.3 | 1.1 | 63.6 | 20.5 | n.d. | 2.5 | 101.2 | $^3$He n.d. | 5.4±1.6 | 3.9±4.3 | 0.15±0.17 | 0.36±0.24[a] | $151_{00} \pm 91_{00}$ | 0.012±0.009 | ~0 |
| Ly4-Cr05 | 0.9±0.3 | 1.8 | 6.7 | 3.3 | 0.9 | 61.7 | 21.6 | n.d. | 1.8 | 97.8 | $^3$He n.d. | 49±15 | 3.1±3.1 | 0.017±0.025 | 1.7±1.1[a] | $29_{000} \pm 16_{000}$ | <0 | <0 |
| Ly4-Cr09 | 1.9±0.2 | 1.9 | 5.8 | 2.6 | 0.9 | 62.6 | 20.8 | n.d. | 4.3 | 98.9 | 0.12±0.09 | 12±1 | $4_0 \pm 90_0$ | <28 | <0.25[a] | <93200 | n.d. | n.d. |
| Ly4-Cr12 | 1.3±0.4 | 2.0 | 7.3 | 1.6 | 0.8 | 48.8 | 41.7 | n.d. | n.d. | 102.2 | 2.2±0.03 | $21_0 \pm 6_0$ | 12.34±0.03 | 0.0314±0.0002 | 8.5±2.5 | 25±0 | 0.17±0.14 | 4.7 |
| Ly4-Cr16 | 1±0.3 | 1.3 | 4.5 | 1.9 | 0.6 | 40.8 | 38.6 | 0.7 | n.d. | 88.4 | 2.31±0.04 | $52_0 \pm 16_0$ | 11.99±0.03 | 0.0308±0.0003 | 16±5 | 33±0.1 | 0.03±0.40 | 0.3 |
| Ly4-Cr17 | 1.3±0.2 | 1.9 | 7.8 | 2.0 | 0.8 | 42.1 | 46.8 | n.d. | n.d. | 101.4 | 2.44±0.04 | $38_0 \pm 6_0$ | 12±0.02 | 0.031±0.0001 | 20±3 | 19±0 | 0.33±0.14 | 9.7 |
| Ly4-Cr21 | 1±0.3 | 2.5 | 4.7 | 2.3 | 0.9 | 58.9 | 28.8 | n.d. | 1.0 | 99.1 | 0.65±0.29 | 5.8±1.7 | 11.93±0.26 | 0.0324±0.0014 | 0.21±0.06 | 27.8±0.8 | 0.031±0.024 | 0.3 |
| Ly4-Cr22 | 0.8±0.2 | 3.4 | 5.9 | 3.1 | 0.8 | 59.1 | 24.8 | n.d. | n.d. | 97.1 | 1.18±0.19 | 16±5 | 11.38±0.14 | 0.0304±0.0006 | 0.73±0.22 | 22.5±0.3 | 0.03±0.03 | 0.1 |
| Ly4-Cr24 | 1.1±0.2 | 3.8 | 2.4 | 1.4 | 0.8 | 54.5 | 31.0 | n.d. | n.d. | 93.9 | $^3$He n.d. | 21±4 | 11.62±0.09 | 0.0305±0.0003 | 0.37±0.07 | 57.3±0.4 | 0.005±0.009 | ~0 |
| Ly4-Cr26 | 4.3±0.2 | 1.9 | 7.2 | 2.4 | 0.9 | 65.6 | 21.0 | n.d. | 7.5 | 106.5 | 0.53±0.07 | 10±0.5 | 5.6±3.1 | 0.23±0.11 | 0.58±0.13[a] | $172_{00} \pm 37_{00}$ | 0.021±0.003 | ~0 |
| Ly4-Cr27 | 7.8±0.1 | 0.6 | 1.1 | 1.1 | 1.4 | 62.0 | 17.0 | n.d. | 10.2 | 93.4 | 0.30±0.03 | 16.8±0.1 | 11.1±2.9 | 0.16±0.04 | 2.0±0.1[a] | $857_0 \pm 56_0$ | 0.023±0.002 | ≡0 |
| Ly4-Cr28 | 0.6±0.2 | 3.0 | 5.2 | 2.2 | 0.9 | 61.4 | 31.1 | n.d. | n.d. | 103.8 | 2.41±0.07 | $18_0 \pm 5_0$ | 11.57±0.05 | 0.0304±0.0002 | 10±3 | 17.1±0.1 | 0.1±0.2 | 2.8 |
| Ly4-Cr29 | 1.4±0.2 | 2.1 | 8.3 | 3.0 | 0.8 | 61.7 | 20.6 | n.d. | 4.3 | 100.8 | $^3$He n.d. | 11±2 | $1_0 \pm 22_0$ | 0.4±6.0 | 0.23±0.41[a] | <127000 | n.d. | n.d. |

*Elemental composition is given in wt%. Values labeled with ([a]), which are given in $10^{-8}$ cm$^3$STP/g. *$^{20}$Ne tripped detector (too high signal). Under "2π", the minimum cosmic-ray exposure ages in the regolith of the L chondrite parent body are listed. All uncertainties are 1σ.*

**Table 2: Grains with short (<1 Ma) cosmic-ray exposure ages**

| Grain | Mass (μg) | $^3$He/$^4$He (×10$^4$) | $^{20}$Ne/$^{22}$Ne | $^{21}$Ne/$^{22}$Ne | $^{21}$Ne$_{cos}$ (×10$^8$) | CRE (Ma) | T$_{PR}$ * (Ma) | Interpretation |
|---|---|---|---|---|---|---|---|---|
| **Sample Ly3** | | | | | | | | |
| Ly3-Cr26 | 11.1±0.2 | 0.199±0.017 | 11.8±4.4 | 0.150±0.057 | 0.0159±0.0024 | 0.25±0.06 | 1.83 | Presence of SW indicates micrometeoritic origin. High mass. |
| Ly3-Cr78 | 8.9±2.7 | 0.382±0.019 | 5.24±3.00 | 0.118±0.065 | 0.0090±0.0031 | 0.14±0.06 | 1.65 | No SW Ne, but clear $^{21}$Ne excess. Ne-HL + E? High mass. |
| **Sample Ly4** | | | | | | | | |
| Ly4-Cr03 | 1.5±0.4 | <0.0893 | 6.30±3.77 | 0.135±0.075 | 0.032±0.015 | 0.51±0.26 | 0.91 | Terrestrial-like He. No SW Ne, but clear $^{21}$Ne excess. Ne-HL + E? |
| Ly4-Cr26 | 4.3±0.2 | 0.526±0.068 | 5.64±3.05 | 0.226±0.115 | 0.021±0.003 | 0.34±0.08 | 1.24 | No SW Ne, but clear $^{21}$Ne excess. Ne-HL + E? High mass. |
| Ly4-Cr27 | 7.8±0.1 | 0.299±0.027 | 11.1±2.9 | 0.155±0.039 | 0.025±0.002 | 0.36±0.08 | 1.52 | Presence of SW indicates micrometeoritic origin. High mass. |

* $T_{PR}$ *is the transfer time for a grain of that size from the asteroid belt to Earth, based on the Poynting–Robertson effect. All uncertainties are 1σ.*